\documentclass{article}
\usepackage{epsfig}
\def\la{\mathrel{\mathpalette\fun <}}
\def\ga{\mathrel{\mathpalette\fun >}}
\def\fun#1#2{\lower3.6pt\vbox{\baselineskip0pt\lineskip.9pt
\ialign{$\mathsurround=0pt#1\hfil##\hfil$\crcr#2\crcr\sim\crcr}}}
\newcommand{\lambar}{\lambda\mkern-10mu\mathchar'26}

\begin{document}
\title{Photon: history, mass, charge
\thanks{Presented at the International Conference on the
Structure and Interactions of the Photon ``Photon 2005: its first
hundred years and the future'', Warsaw University,
31.08.--04.09.2005} }
\author{L.B. Okun
\\ \small 117218 ITEP, Moscow, Russia }
\maketitle
\begin{abstract}
The talk consists of three parts. ``History'' briefly describes
the emergence and evolution of the concept of photon during the
first two decades of the 20th century. ``Mass'' gives a short
review of the literature on the upper limit of the photon's mass.
``Charge'' is a critical discussion of the existing interpretation
of searches for photon charge. Schemes, in which all photons are
charged, are grossly inconsistent. A model with three kinds of
photons (positive, negative and neutral) seems at first sight to
be more consistent, but turns out to have its own serious
problems.
\end{abstract}

\section{History}

The idea that light consists of rapidly moving particles can be
traced from the writings of ancient authors to Descartes and
Newton. The wave theory of light was put forward by Huyghens and
was later decisively proved to be correct through discovery of
interference and diffraction by Young and Fresnel. Maxwell's
theory of light as electromagnetic waves was one of the greatest
achievements of the 19th century.

The history of the photon in the 20th century started in 1901 with
the formula by Planck for radiation of a black body and
introduction of what was called later the quantum of action $h$
\cite{1}. In 1902 Lenard discovered that energy of electrons in
photoeffect does not depend on the intensity of light, while it
depends on the wavelength of the latter \cite{2}.

In his fundamental article ``On an euristic point of view
concerning the production and transformation of light'' published
in 1905 Einstein pointed out that the discovery of Lenard meant
that energy of light is distributed in space not uniformly, but in
a form of localized light quanta \cite{3}. He has shown that all
experiments related to the black body radiation, photoluminescence
and production of cathode rays by ultraviolet light can be
explained by the quanta of light.

The proof that Einstein's light quanta behave as particles,
carrying not only energy, but also momentum, was given in 1923 in
the experiments by Compton on scattering of X-rays on electrons
\cite{4}.

The term ``photon'' for particles of light was coined by Lewis in
1926 in an article ``The conservation of photons'' \cite{5}. His
notion of a photon was different from the notion we use today. He
considered photons to be ``atoms'' of light, which analogously to
the ordinary atoms are conserved.

The term ``photon'' was quickly accepted by physics community. The
fifth Solvay Council of physics, which took place on October
24-29, 1927, had the name ``Electrons and Photons'' \cite{6}. The
term ``photon'' in its present meaning was first used in the talk
by Compton at this meeting  (see ref. \cite{6}, p. 55).

In his talk Compton used the term ``photon'' as if it existed
since  1905; thus on page 62 of ref. \cite{6} one can read: ``It
is known that the hypothesis of photons was introduced by Einstein
in order to explain the photo-electric effect''. On the other
hand, on page 57 one can read:

``When speaking of this unit of radiation, I would use the name
``photon'' suggested recently by G.N.~Lewis (Nature, 18 December,
1926). ... it has the advantage of being brief without implying
any relation with mechanics of quanta, more general, or the
quantum theorie of atomic structure''.

The Proceedings \cite{6} open with an obituary of H.A.~Lorentz who
passed away in February 1928, a few months after the Fifth Solvay
meeting, in which Lorentz actively participated.

The speakers at the meeting were:

W.-L.~Bragg, The intensity of reflected X-rays, pp. 1-44;

A.H.~Compton, Discordances  between the experiment and the
electromagnetic theory of radiation, pp. 55-86;

L.~de Broglie, The new dynamics of quanta, pp. 105-133;

M.~Born and W.~Heisenberg, The mechanics of quanta, pp. 143-182;

E.~Schr\"{o}dinger, The mechanics of waves, pp. 185-207;

N.~Bohr, The postulate of quanta and the development of
atomistics, pp. 215-248.

Each of the talks was followed by a detailed discussion.
Participated Bohr, Born, Brillouin, de Broglie, Compton, Dirac, De
Donder, Ehrenfest, Fowler, Heisenberg, Kramers, Langmuir,
Langevin, Lorentz, Pauli, Richardson, Schr\"{o}dinger.

Einstein took part only in the ``General discussion of the new
ideas'', expressed during the meeting. The discussion (pp.
248-289) was presented in three sections: 1. Causality,
Determinisme, Probability; 2. Photons; 3. Photons and Electrons.

Einstein spoke in the first section (pp. 253-256) and asked a
question during the second section (p. 266). He considered a
screen with a small hole in it and a spherical layer of
photoemulsion of large radius behind it. Electrons fall on the
screen as De Broglie - Schr\"{o}dinger plane waves normal to it
and reach the emulsion as spherical waves. Einstein discussed the
two possible interpretations of this thought experiment: purely
statistical and purely deterministic. The term ``photon'' was not
used in his remarks.

The term ``photon'' was again used by Compton on December 12,
1927, this time without any reference to Lewis, in Compton's Nobel
lecture ``X-rays as a branch of optics'' \cite{7}. On page 186 one
can read:

``An X-ray photon is deflected through an angle $\phi$ by an
electron, which in turn recoils at an angle $\theta$, taking a
part of the energy of photon''.

Further on page 187:

``... recoil electrons are in accord with the predictions of the
photon theory''.

The year 1927 marked the end of the history of emergence of the
concept of photon. A few years later Dirac opened a new chapter in
Physics by establishing Quantum Electrodynamics.

As for Einstein, he wrote in 1951:

``All these fifty years of pondering have not brought me any
closer to answering the question, What are light quanta?''
\cite{pais}.

\section{Mass}

The problem of the upper limit on the mass of the photon was
raised at ITEP by Isaak Yakovlevich Pomeranchuk (20.05.1913 --
14.12.1966) in autumn of 1966, a few months before he lost his
fight against cancer. He put this question to his former students:
Igor Yuryevich Kobzarev (15.10.1932 -- 20.01.1991) and myself.
First we wrote a draft of a short research note, but then after a
thorough search we discovered that most of our considerations had
been already addressed in the literature by de~Broglie \cite{9},
\cite{10} (see also \cite{11}, \cite{12}), Schr\"{o}dinger
\cite{13}, \cite{14}, Bass and Schr\"{o}dinger \cite{15} and by
Gintsburg \cite{16}.

In particular, de~Broglie \cite{10} noticed that photon mass would
lead to a faster speed of violet light than that of the red one.
He concluded that during the eclipse in double star system the
color of the appearing star would change from violet to red. He
also considered the dispersion of radiowaves.

Schr\"{o}dinger \cite{13,14} pointed out that magnetic field of
the Earth would be exponentially cut off at distances of the order
of the photon Compton wave length $\lambar_\gamma = 1/m_\gamma$.
From the observed altitude of auroras he concluded that
$\lambar_\gamma
> 10^4$ km.

Gintsburg \cite{16} corrected the limit of Schr\"{o}dinger and
suggested that measurements of the magnetic field of Jupiter could
improve the limit to $\lambar_\gamma \sim 10^6$ km. He also was
the first to consider how the mass of the photon would influence
the magnetohydrodynamic waves in plasma.

These results discouraged us from publishing an original article.

From the beginning of 1968 a special issue of Uspekhi Fizicheskikh
Nauk was under preparation to mark 50 years of this review
journal. Kobzarev and I were invited to publish our paper on
photon as a review.

In this short review \cite{17} we corrected the estimates by
de~Broglie and Schr\"{o}dinger. The former estimate was
invalidated due to dispersion of light in the atmospheres of stars
(we found that this effect was considered by Lebedev in 1908
\cite{18}). Ref. \cite{18} was the paper, which closed the
discussion of colour variation in binary stars. The effect was
discovered by Belopolskii \cite{19}, Nordmann \cite{20} and
Tikhoff \cite{21} and interpreted by them as dispersion of light
in the interstellar free space. The observed minimum in red light
preceded that in violet light from the variable binary stars by a
few minutes. (Note that for a massive photon the violet light
should be faster, not slower!) Lebedev\footnote{Petr Nikolaevich
Lebedev (1866--1912) is  famous by his experimental discovery of
pressure of light.} rejected this interpretation and explained the
effect by the difference of pressure in the atmospheres of two
stars \cite{22, 23}.

We also found that a much better limit could be extracted from the
measurements by Mandelstam \cite{24} of radiowave dispersion in
the atmosphere of the Earth.

As for the limit by Schr\"{o}dinger we conservatively  extended it
to 30~000 km by using the data from review by Bierman \cite{25},
though these data (from rockets and satellites) indicated the
spread of the geomagnetic field to 60~000 km and even to 100~000
km.

In addition to magnetic field we have interpreted in terms of
$\lambar_\gamma$ the experiment by Plimpton and Lawton \cite{26},
testing the absence of Coulomb field in the space between two
concentric spheres, and derived $\lambar_\gamma < 10$ km. (The
deviation from Coulomb law was parametrized in ref. \cite{26} by
$1/r^{2+\varepsilon}$.)

We also discussed why the longitudinal photons do not manifest
themselves in the black body radiation, a subject considered by
Bass and Schr\"{o}dinger \cite{15}.

Our review \cite{17} appeared in May 1968.

Two months later Physical Review Letters received and in August
published a paper by Goldhaber and Nietto \cite{27} ``New
geomagnetic limit on the mass of the photon''. Their geomagnetic
limit was about 90~000 km. They derived $\lambar_\gamma < 10$ km
from reference \cite{26} and reconsidered the geomagnetic
estimates by Gintsburg \cite{16}.

Three years later Goldhaber and Nietto published an extensive
review \cite{28} with about 100 references. The review by Byrne
\cite{29} published in 1977 has about 40 references. The latest
review by Tu, Luo and Gillies \cite{30} published in 2005 has
about 200 references.

It is impossible to comment on all these hundreds of papers in a
short review. One has to make selection.

Since 1992 the selected references on the photon mass are cited by
the Particle Data Group (PDG) in biennial Reviews of Particle
properties \cite{31} - \cite{37}. The best cited limits (in eV)
were chosen by PDG:

\bigskip
\noindent 1992: $3 \cdot 10^{-27}$, Chibisov \cite{38}, galactic
magnetic field. \\ 1994: $3 \cdot 10^{-27}$, Chibisov \cite{38},
galactic magnetic field. \\ 1996: $6 \cdot 10^{-16}$, Davis et al.
\cite{39}, Jupiter magnetic field. \\ 1998: $2 \cdot 10^{-16}$,
Lakes \cite{40}, torque on toroid balance. \\ 2000: $2 \cdot
10^{-16}$, Lakes \cite{40}, torque on toroid balance. \\ 2002: $2
\cdot 10^{-16}$, Lakes \cite{40}, torque on toroid balance. \\
2004: $6 \cdot 10^{-17}$, Ryutov \cite{41}, magnetohydrodynamics
of solar wind (MHD).

If $c$ is the unit of velocity and $\hbar$ is the unit of action,
then 1 eV $= 1.78 \cdot 10^{-33}$ g., 1 eV $= (1.97 \cdot 10^{-10}
\; {\rm km})^{-1}$.

Chibisov \cite{38} considered the conditions of equilibrium of
magnetic field in the smaller Magellanic cloud by applying virial
theorem. This gave $\lambar_\gamma \la l$, where $l$ is the size
of the cloud ($l \simeq 3 \; {\rm k p c} = 3 \cdot 3.08 \cdot
10^{16} \; {\rm km} \approx 10^{17} \; {\rm km} = 10^{22} \; {\rm
cm}$). It is not clear how reliable is this approach.

Davis et al. \cite{39} used the ``Jupiter suggestion'' of
Gintsburg \cite{16} and the new Pioneer-10 data on the magnetic
field of Jupiter.

A novel idea was put forward and realized by Lakes \cite{40}. He
exploited the fact that the term $m_\gamma ^2 A^2$ in Lagrangian
breaks the gauge invariance of Maxwell's electrodynamics. In
Lorenz\footnote{Quite often the Lorenz gauge is erroneously
ascribed to Lorentz (for clarification and for earlier references
see ref. \cite{42}).} gauge one has the Maxwell-Proca equation. As
a result the vector potential {\bf A} becomes observable. Lakes
performed an experiment with a toroid Cavendish balance to search
for the torque $m_\gamma ^2 A$ produced by the ambient vector
potential $A$.

The experiment \cite{40} disclosed that $A m_\gamma ^2 < 2 \cdot
10^{-9} \; {\rm T m}/{\rm m}^2$. If the cosmic vector potential
$A$ is $10^{12}$ Tm, then $\lambar_\gamma = m_\gamma^{-1} \ga 2
\cdot 10^{10}$ m. This limit has been improved by other authors
(see ref. \cite{30}). However the estimate of the value of cosmic
potential $A$ is not reliable enough.

Ryutov \cite{41} developed the idea of Gintsburg \cite{34} and
first derived a selfconsistent and complete set of MHD equations
accounting for finite photon mass. He did not put a new limit on
the photon mass, but mentioned a possible way of improving it by
the analysis of the sector structure of the Solar wind. In
particular he noticed that the limit $6 \cdot 10^{-16}$ eV,
considered in 1996 by PDG as the best one should be reduced by
approximately an order of magnitude. This is the origin of the PDG
best number in 2004.

\section{Charge}

There exist about a dozen of  papers \cite{43} - \cite{52}
questioning the neutrality of photons and setting an upper limit
on their charge. In all of them the upper limit follows from the
non-observation of any action  of external static electric or
magnetic fields on photon's charge, while the fact that these
fields themselves are ``built from photons'' is ignored. As a
result all those papers \cite{43} - \cite{52} lack a
self-consistent phenomenological basis. But without such a basis
any interpretation of experimental data is meaningless.

In fact the authors \cite{43} - \cite{52} implicitly assumed that
all photons are either neutral as in ordinary QED, or all are
charged. It is obvious that the latter assumption is impossible to
reconcile with the existence of classical static electric or
magnetic fields. Hence the best upper limit on the value of photon
charge presented by the Particle Data Group \cite{45} seems to be
meaningless.

It is clear, that for a more consistent interpretation of searches
\cite{43} - \cite{52} both types of photons are necessary: charged
and neutral. In such a scheme classical electric and magnetic
fields are built from the latter. Hence the scattering of all
charged particles (including the charged photons)  by these fields
occurs due to absorption of virtual neutral photons. Charge is
conserved in this processes. (The failure of theoretical attempts
to violate the conservation of electric charge was analyzed in
references \cite{53,54}.)

However a scheme with both charged and neutral photons is also not
without serious problems. One of them is the catastrophic infrared
emission of neutral photons by massless charged ones. The other
problems are connected with the emission and absorption of charged
photons by ordinary charged particles, say, electrons.

Conservation of charge calls in this case for the existence of a
twin electron with charge $e - e'$, where $e'$ is the charge of
the emitted charged  photon, which is assumed  to be  much smaller
than $e$. The mass of the twin must be much larger than the mass
of the electron in order to avoid contradiction with data on
atomic, nuclear, and high energy physics.

One might consider the three photons with charges  $+e'$, $-e'$, 0
as an SU(2) Yang--Mills triplet, while the electron with charge
$e$ and its twin with charge $e - e'$ as an SU(2) doublet. The
SU(2) symmetry requires mass degeneracy of particles belonging to
the same multiplet. However even in this degenerate case it is
impossible to accommodate the inequality $e'/e \ll 1$ in a scheme
without ``astronomically huge'' Higgs multiplets. The situation is
further aggravated by the breaking of SU(2) gauge symmetry,
responsible for the difference of masses of particles and their
twins.
\bigskip

\section*{Acknowledgments}

I am grateful to V.P.~Vizgin, K.A.~Tomilin and A.D. Sukhanov for
helpful discussions on the history of the concept of photon.

I am grateful to A. Buras, V. Fadin, and P. Zerwas for drawing my
attention to the articles on hypothetical charge of the photon and
to G. Cocconi, G. Giudice, and M. Vysotsky  for valuable remarks.

I am grateful to M. Krawczyk for wonderful hospitality.

The work was supported by the grant of the Russian ministry of
education and science No. 2328.2003.2.


\begin{thebibliography}{99}
\bibitem{1}
M. Planck, Ann. Phys. {\bf 4} (1901) 561.
\bibitem{2}
P. Lenard, Ann. Phys. {\bf 8} (1902) 169.
\bibitem{3}
A. Einstein, Ann. Phys. {\bf 17} (1905) 132.
\bibitem{4}
A.H. Compton, Phys. Rev. {\bf 22} (1923) 409.
\bibitem{5}
G.N. Lewis, Nature, No. 2981, vol. 118 (December 18, 1926) 874.
\bibitem{6}
{\it Electrons et Photons.} Rapports et discussions du cinquiem
conseil de physique tenu a Bruxelles du 24 au 29 octobre 1927 sous
les auspices de l'Institut International de Physique Solvay.
Paris, Gauthier-Villars et c$^{\rm ie}$ Editeurs, 1928.
\bibitem{7}
A.H. Compton, {\it in} ``Nobel Lectures. Physics. 1901-1995''
CD-ROM, World Scientific, 1998.
\bibitem{pais}
Abraham Pais, {\it 'Subtle in the Lord ...' The Science and the
Life of Albert Einstein}, Section 19f, p. 382, Oxford, 1982.
\bibitem{9}
L. de Broglie, Phil. Mag. {\bf 47} (1924) 446.
\bibitem{10}
L. de Broglie,  {\it La mechanique  ondulatoire du photon. Une
nouvelle theorie de la lumiere, tome premier}, Paris, 1940, pp.
39, 40.
\bibitem{11}
L. de Broglie, {\it La mechanique ondulatoire du photon et theorie
quantique de champs}, Paris, 1949.
\bibitem{12}
L. de Broglie, {\it La theorie generale des particule \'a Spin},
Paris, 1943, p. 191.
\bibitem{13}
E. Shr\"{o}dinger, Proc. Roy. Irish Acad. {\bf A49} (1943) 43.
\bibitem{14}
E. Shr\"{o}dinger, Proc. Roy. Irish Acad. {\bf A49} (1943) 135.
\bibitem{15}
L. Bass, E. Shr\"{o}dinger, Proc. Roy. Soc. {\bf A232} (1955) 1.
\bibitem{16}
M.A. Gintsburg, Astronom. Zhurnal {\bf 40} (1963) 703-709 (in
Russian); \\ M.A. Gintsburg, Sov. Astron. -- AJ {\bf 7} (1964)
536-540 (English translation).
\bibitem{17}
I.Yu. Kobzarev, L.B. Okun, Uspekhi Fiz. Nauk {\bf 95} (1968)
131-137; \\ I.Yu. Kobzarev, L.B. Okun, Sov. Phys. Usp. {\bf 11}
(1968) 338-341.
\bibitem{18}
P.N. Lebedev, Sobranie sochineniy (Collected papers), M., Izd. AN
SSSR, 1963, paper 33 (in Russian).
\bibitem{19}
A.A. Belopolskii, Izvestia Imper. Akademii Nauk {\bf 21} (1904)
153 (in Russian).
\bibitem{20}
C. Normann, Comptes Rendus {\bf 146} (1908) 266, 383.
\bibitem{21}
G.A. Tikhoff, Comptes Rendus {\bf 146} (1908) 570.
\bibitem{22}
P.N. Lebedev, Izv. Imp. AN {\bf 24} (1906) 93 (in Rusian).
\bibitem{23}
P.N. Lebedev, Comptes Rendus {\bf 146} (1908) 1254; {\bf 147}
(1908) 515.
\bibitem{24}
L.I. Mandelshtam, {\it Polnoe sobranie trudov (Complete works)},
Izv. AN SSSR, 1947, v. 2, pp. 277-305; v. 3, p. 238 (in Russian).
\bibitem{25}
L. Bierman, Sitzber. Bayerische Akad. Wissenschaften {\bf 37}
(1965).
\bibitem{26}
S.J. Plimpton, W.E. Lawton, Phys. Rev. {\bf 50} (1936) 1066.
\bibitem{27}
A.S. Goldhaber, M.M. Nieto, Phys. Rev. Lett. {\bf 21} (1968) 567.
\bibitem{28}
A.S. Goldhaber, M.M. Nieto, Rev. Mod. Phys. {\bf 43} (1971) 277.
\bibitem{29}
J.C. Byrne, Astrophysics and Space Science {\bf 46} (1977)
115-132.
\bibitem{30}
L-C. Tu, J. Luo, G.T. Gillies, Rep. Prog. Phys. {\bf 68} (2005)
77-130.
\bibitem{31}
PDG, Phys. Rev. D {\bf 45} (1992) V.1.
\bibitem{32}
PDG, Phys. Rev. D {\bf 50} (1994) 1351.
\bibitem{33}
PDG, Phys. Rev. D {\bf 54} (1996) 207.
\bibitem{34}
PDG, Eur. Phys. J. {\bf 3} (1998) 223.
\bibitem{35}
PDG, Eur. Phys. J. {\bf 15} (2000) 249.
\bibitem{36}
PDG, Phys. Rev. D {\bf 66} (2002) 281.
\bibitem{37}
PDG, Phys. Lett. B {\bf 592} (2004) 335.
\bibitem{38}
G.V. Chibisov, Usp. Fiz. Nauk {\bf 119} (1976) 551-5 (in Russian);
Sov. Phys. Usp. {\bf 19} (1976) 624-6 (English translation).
\bibitem{39}
L. Davis, A.S. Goldhaber, M.M. Nieto, Phys. Rev. {\bf 35} (1975)
1402-5.
\bibitem{40}
R. Lakes, Phys. Rev. Lett. {\bf 80} (1998) 1826-9.
\bibitem{41}
D.D. Ryutov, Plasma Phys. Control Fusion {\bf 39} (1997) A73-A82.
\bibitem{42}
J.D. Jackson, L.B. Okun, Rev. Mod. Phys. {\bf 73} (2001) 663-680.
\bibitem{43}
L. Grodzins, D. Engelberg, W. Bertozzi, Bull.  Am. Phys. Soc. {\bf
6} (1961) 63.
\bibitem{44}
R. W.  Stover, T.I.  Moran,  and J.W. Trischka, Phys. Rev. {\bf
164} (1967) 1599.
\bibitem{45}
PDG, ``Review of Particle Physics'', Phys. Lett. B {\bf 592}
(2004) 31, 335.
\bibitem{46}
G. Cocconi, Phys. Lett. B {\bf 206} (1988) 705.
\bibitem{47}
G. Raffelt, Phys. Rev D {\bf 50} (1994) 7729; hep - ph/0411398.
\bibitem{48}
G.  Cocconi, Am. J. Phys. {\bf 60} (1992) 750.
\bibitem{49}
V.V. Kobyshev and S.B. Popov, hep- ph/0411398.
\bibitem{50}
C. Sivaram, Am. J. Phys. {\bf 63} (1994) 1473.
\bibitem{51}
C. Caprini and P.G. Ferreira , JCAP {\bf 0502} (2005) 006;
hep-ph/0310066.
\bibitem{52}
Y.K. Semertzidis, G.T. Danby, D.M. Lazarus, Phys. Rev. D {\bf 67}
(2003) 017701.
\bibitem{53}
L.B. Okun and Ya.B. Zeldovich, Phys. Lett. B {\bf 78} (1978) 597.
\bibitem{54}
M.B. Voloshin and L.B. Okun, Pis'ma ZhETF {\bf 28} (1978) 156 (in
Russian); JETP Lett. {\bf 28} (1978)145.

\end{thebibliography}
\end{document}